\documentstyle[epsfig,aps,prl,epsf,multicol]{revtex}

\begin{document}


\hsize\textwidth\columnwidth\hsize
\csname@twocolumnfalse\endcsname

\title{Dynamic melting of confined vortex matter}

\author{R.~Besseling, N.~Kokubo and P.H.~Kes}

\address{Kamerlingh Onnes Laboratorium, Leiden
University, P.O. Box 9504, 2300 RA Leiden, the Netherlands.}
\date{\today}
\maketitle

\begin{abstract}
We study {\em dynamic} melting of confined vortex matter moving in
disordered, mesoscopic channels by mode-locking experiments. The
dynamic melting transition, characterized by a collapse of the
mode-locking effect, strongly depends on the frequency, i.e. on
the average velocity of the vortices. The associated dynamic
ordering velocity diverges upon approaching the equilibrium
melting line $T_{m,e}(B)$ as $v_c \sim (T_{m,e}-T)^{-1}$. The data
provide the first direct evidence for velocity dependent melting
and show that the phenomenon also takes place in a system under
disordered confinement. 
\end{abstract}
\begin{multicols}{2}
\narrowtext \noindent

An intriguing aspect of periodic media driven through a (random)
pinning potential, e.g. vortex lattice (VL's) in superconductors
\cite{Blatterbible}, charge density waves (CDW's) \cite{Gruner},
or conventional sliding solids \cite{Perssonbook}, is the
possibility of a dynamic ordering (DO) transition
\cite{BhattaShi,Koshelevrecryst}: while at small velocity pinning
may disrupt the lattice, causing liquid-like flow
\cite{Jensen_filam_PRL88} around pinned islands, at large velocity
the influence of pinning diminishes and the elastic interactions
dominate, favoring a crystalline structure.

The DO phenomenon becomes particularly interesting close to the
{\it thermodynamic} melting point of the medium, since in addition
to fluctuations due to pinning also thermal fluctuations become
relevant \cite{Koshelevrecryst,Balents}. Their combined effect was
first studied by Koshelev and Vinokur \cite{Koshelevrecryst} for
2D VL's. They introduced the concept of a {\it shaking
temperature} $T_{sh}$, characterizing the fluctuations in the
moving frame due to the quenched disorder. This shaking diminishes
with increasing velocity as $k_BT_{sh}=\Gamma_{p,v}/v$
($\Gamma_{p,v}$ reflects the pinning strength and viscous
damping). Fluid-like, incoherent motion sets in when the effective
'temperature' $T+T_{sh}$ equals the equilibrium melting
temperature $T_{m,e}$ of the pure system. Hence, for $T<T_{m,e}$
there exists a recrystallization velocity
\begin{equation}
v_c=\Gamma_{p,v}/[k_B(T_{m,e}-T)], \label{DOvsTformula}
\end{equation}
which diverges when temperature approaches $T_{m,e}$. Indications
for this scenario were found in experiments on superconducting
films \cite{HellerGeers} by identifying an inflection point in the
current-voltage ($IV$) curves with DO. This identification seemed
justified from simulations \cite{RyuPRL96} but more direct
experimental evidence for the phenomenology in
\cite{Koshelevrecryst} is still lacking. Simulations of a density
gradient driven system \cite{Basslergradientsim} in fact suggest
that the inflection point can also be due to changes in
large-scale flow morphology while recent experiments on crystals
\cite{PaltielIV} have shown that for a $3$D VL the inflection
point is due to macroscopic coexistence of two phases and not due
to {\it microscopic} DO.

In this Letter we study DO of {\it confined} vortex matter flowing
in disordered mesoscopic channels using a completely different
dynamic probe. In \cite{KokuboPRL02} we showed how a mode-locking
(ML) technique can be used to explore the flow configuration in
the channels in presence of strong disorder from the vortex arrays
in the channel edges. The ML phenomenon occurs due to coupling
between, on the one hand, lattice modes of frequency $f_{int}=q
v/a$, with $q$ an integer and $a$ the lattice periodicity, which
may exist in an array that moves coherently with velocity $v$,
and, on the other hand, a superimposed rf-drive of frequency $f$
at an integer fraction $1/p$ of $f_{int}$ \cite{Fiory}. This
coupling produces plateaus in the dc-transport ($IV$) curves or
sharp peaks in the differential conductance ($dI/dV$) when
$v=(p/q) f a$. However, these peaks are reduced and can eventually
vanish due to incoherent velocity fluctuations in the moving
array, arising from both thermal and quenched disorder
\cite{HarrisPRL95,KoltonshapPRL01}. Thus, the collapse of the ML
peak marks the transition from coherent to incoherent flow. Using
this criterion we provide for the first time conclusive evidence
for a velocity dependent melting transition at $v_c$ \cite{fn} and
probe the divergence of $v_c$ upon approaching the static phase
boundary.

The experiments are performed on a superconducting double layer of
weak pinning amorphous (a-)Nb$_{1-x}$Ge$_x$ ($x \simeq0.3$, $550$
nm thickness, $T_c=2.68$ K, normal resistivity $\rho_n=2
\mu\Omega$m) and strong pinning NbN ($50$ nm thickness) on top,
containing $N_{ch}(\simeq 200)$ straight channels of width $w=230$
nm etched to a depth of $300$ nm \cite{TheunissenPRL}. Vortices in
the channels are confined between channel edges (CE's) consisting
of pinned, disordered VL's which impose both a periodic and random
potential on the confined vortices via their mutual shear
interaction, characterized by a shear modulus $c_{66}$. This
potential determines both the threshold force density $F_p\sim
c_{66}/w$ and the dynamics of vortices in the channel
\cite{KokuboPRL02,BesselingEPL03,BesselingPhDthesis}. We measured
dc and dc-rf transport versus magnetic field and temperature using
a four probe configuration. The sample was immersed in superfluid
$^4$He. The frequency of the applied rf current was in the range
$1-200$ MHz, its amplitude $I_{rf}$ could be as large as $2$ mA
\cite{fnNbNpinfreq}.

Due to the confinement, the ML condition for channel flow attains
a particularly useful form: although the array in the channel may
be frustrated, meaning that the average longitudinal periodicity
$a\neq a_0$ and the row spacing $b\neq b_0$
($a_0=2b_0/\sqrt{3}\simeq 1.075\sqrt{\Phi_0/B}$ are the
equilibrium values), the {\em voltage} at which the main
interference ($p=q=1$) occurs is simply given by
\cite{KokuboPRL02}
\begin{eqnarray}
V_{1,1} = \Phi_0 f n N_{ch}, \label{V1relation}
\end{eqnarray}
with $n$ the number of moving rows in each channel.

\begin{figure}
\epsfig{file=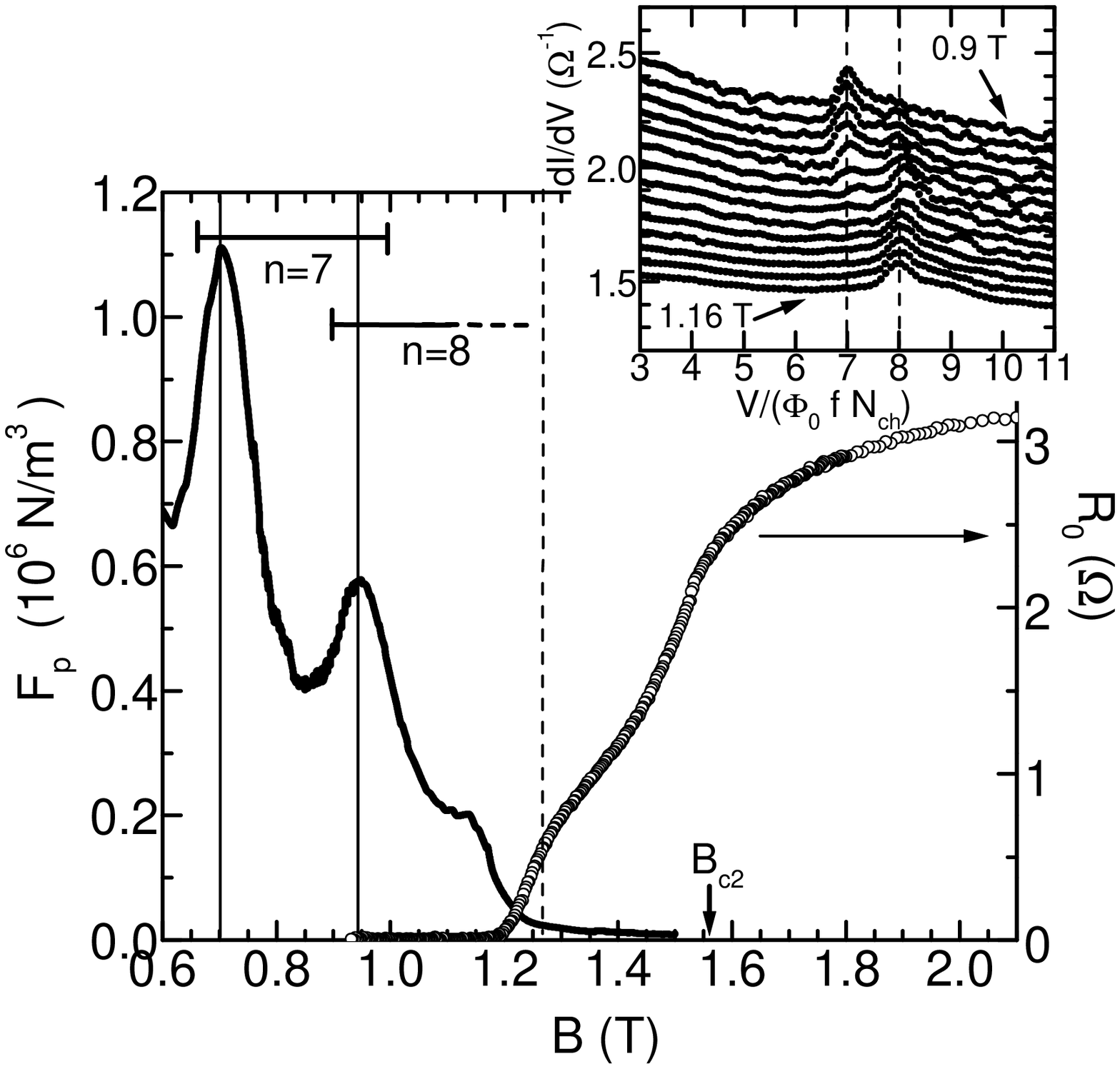, width=7.5cm} \vspace{0.2cm} \caption{Shear
force density (determined using a velocity criterion
$v/a_0\simeq1$ MHz) versus field at a temperature $T=1.94$ K.
Vertical lines indicate the field of the transition from $n=6
\rightarrow 7$ and $n=7 \rightarrow 8$, dashed vertical line marks
the field where the $n=8 \rightarrow 9$ transition is expected.
Open symbols (right axis): field dependence of the zero bias
resistance $R_0$. Inset: $dI/dV$ versus normalized voltage in
presence of $85$ MHz rf-currents for $B=0.9-1.16$ T (steps of $20$
mT). The data illustrate the $n=7 \rightarrow 8$ transition in the
ML signal.} \label{fig1}
\end{figure}

We first discuss the typical behavior in the solid phase where
matching effects between the confined array and the channel width
dominate the behavior. In Fig. \ref{fig1} the thick line shows the
dc-depinning force density $F_p=J_pB$ ($I_{rf}=0$). For fields
$B\lesssim1.1$ T two oscillations in $F_p$ are seen. The inset
shows $dI/dV$ curves in presence of an $85$ MHz rf-current
($I_{rf}=0.53$ mA) versus voltage for fields $0.9$ T$\leq B\leq
1.16$ T. The data for $0.9$ T exhibits a peak in $dI/dV$
corresponding to ML of $7$ vortex rows in each channel. For larger
fields the amplitude of this peak decays while another peak,
corresponding to $8$ rows, appears. The field region where these
peaks coexist ($0.92$ T $\lesssim$ B $\lesssim 1$ T) evidently
corresponds to the situation of maximum mismatch between the
natural width $nb_0$ of $n=7,8$ vortex rows and the effective
channel width (estimated as $w_{eff}=315$ nm). In the same field
regime $F_p$ exhibits a maximum. As shown in \cite{KokuboPRL02},
this maximum at mismatch is caused by jamming of the flow at
locations in the channel where the number of rows switches from
$n$ to $n \pm 1$. The motion there is partially blocked by
dislocations with Burgers vectors that are almost perpendicular to
the flow direction. The structural disorder of the fixed VL in the
CE's is responsible for this phenomenon.

We now turn to the behavior in larger magnetic fields, $B\gtrsim
1.1$ T. At $B\simeq 1.15$ T, $F_p$ shows a rapid decrease after a
small upturn. This drop coincides with the onset of a measurable
zero-bias resistance $R_0$ (displayed on the right axis). These
two distinct features reflect the loss of shear rigidity of the
vortex configuration inside the channel, indicating a transition
to a confined vortex liquid \cite{TheunissenPRL}. Note that these
features occur well below the magnetic field at which the
transition from $n=8$ to $9$ vortex rows is expected from the
condition $8.5b_0=w_{eff}$, namely $B_{8,9}=1.27$ T. In addition,
neither $F_p$ nor $R_0$ show any particular features at $B_{8,9}$,
which shows that the liquid is insensitive to (mis)matching
effects.

\begin{figure}
\epsfig{file=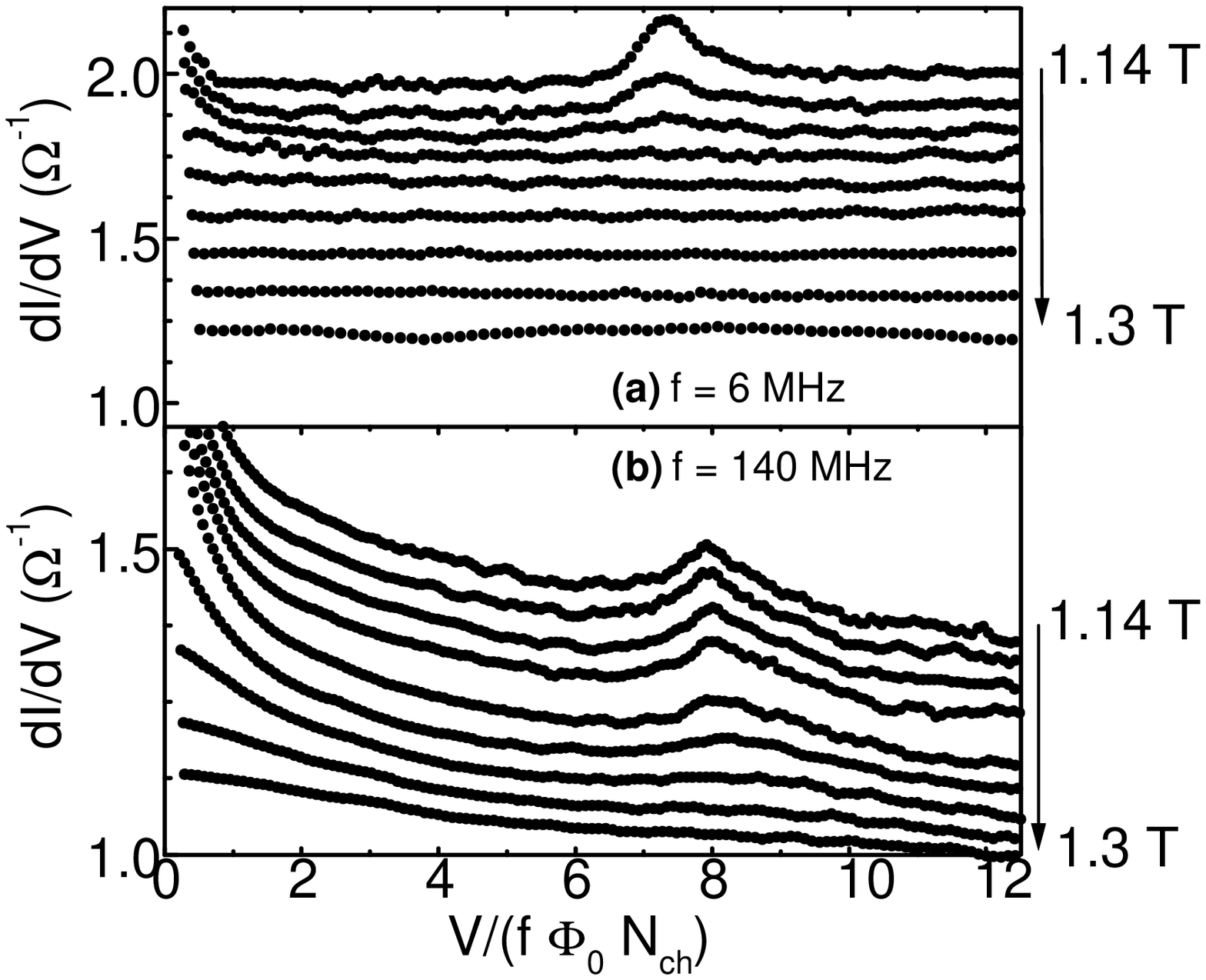, width=7cm} \vspace{0.2cm}
\caption{Differential conductance versus reduced voltage at
$T=1.94$ K in fields ranging from $1.14$ T (topcurve) to $1.3$ T
($20$ mT steps) for (a) rf-current of $6$ MHz ($I_{rf}=0.09$ mA)
and (b) $140$ MHz and $I_{rf}=0.94$ mA.} \label{fig2}
\end{figure}

The ML experiments provide important new information regarding the
coherence and the shear rigidity of the {\it moving} array. Figure
\ref{fig2}(a) shows $dI/dV$ data in the field range $1.14-1.3$ T
plotted versus the normalized voltage for a $6$ MHz rf-current.
The ML peak in the upper curve ($B=1.14$ T) corresponds to the
coherent motion of $8$ rows. For larger field the peak amplitude
drops rapidly and vanishes at $B\simeq 1.2$ T. Above $1.2$ T the
$dI/dV$ versus $V$ curves remain featureless. The vanishing of the
ML signal marks the transition from coherent motion of a vortex
solid to incoherent 'liquid' motion. A similar interpretation of
this phenomenon was suggested in \cite{HarrisPRL95} with respect
to the VL melting transition in YBCO.

In Fig. \ref{fig2}(b) a $dI/dV$ data set for the same field range
is shown, but now for a $140$ MHz rf-current. The average vortex
velocity at mode-locking, $v=f a $ is therefore over $20$ times
larger than in Fig. \ref{fig2}(a). Again a clear ML peak is
observed at $B=1.14$ T, but in contrast to the $6$ MHz data in (a)
the field range in which ML takes place is considerably larger,
extending up to $B\simeq 1.3$ T. The nature of the moving medium
thus depends strongly on its velocity. For example: at $B=1.24$ T
one observes 'liquid' motion for $f=6$ MHz but coherent motion at
$140$ Mhz. Interestingly, in the early work of Fiory \cite{Fiory}
it was already mentioned that the vanishing of the ML signal
shifts to larger fields when measured at larger frequency.

\begin{figure}
\epsfig{file=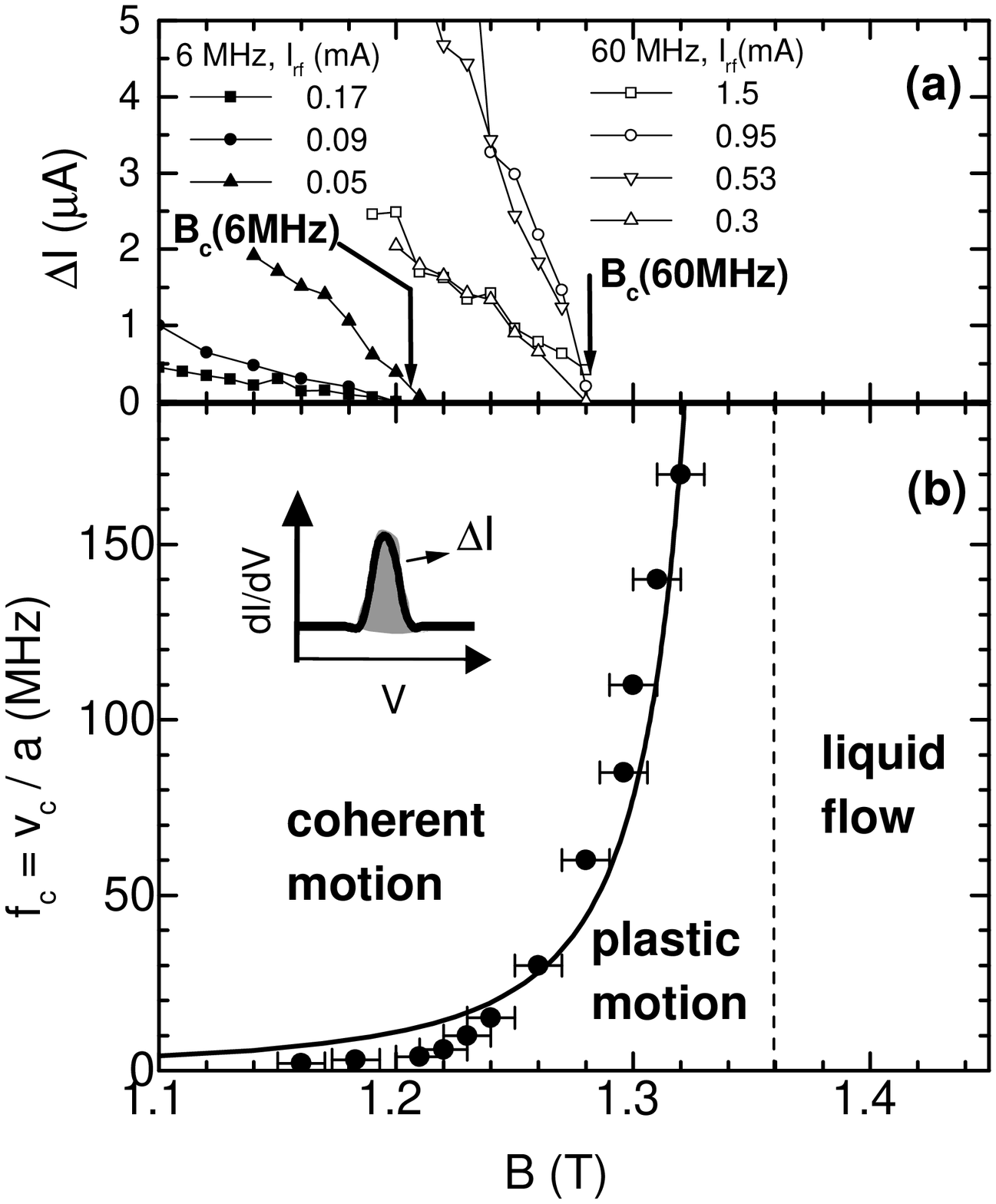, width=6.8cm} \vspace{0.2cm} \caption{(a) ML
step width $\Delta I$, as defined in the inset to (b), versus
field for $f=6$ MHz and $f=60$ MHz and several values of $I_{rf}$.
Arrows indicate the dynamic field $B_c(f)$. (b) Data points: $f_c$
versus $B$. Drawn line is a fit according to $f_c\sim
(B_{m,e}-B)^{-2}$. The dashed lined marks $B_{m,e}=1.36$
T.}\label{fig3}
\end{figure}

For a proper determination of the collapse of the ML signal we
should take into account its dependence on $I_{rf}$ (see e.g.
\cite{ThorneMLampPRB87}). Therefore the ML step width $\Delta I$
(defined in the inset to Fig. \ref{fig3}(b)) was determined as
function of field for various rf-amplitudes and frequencies. The
result is shown in Fig. \ref{fig3}(a) for $6$ MHz and $60$ Mhz. As
observed, $\Delta I$ vanishes linearly with $B$ at a {\it dynamic
melting field} $B_c(f)$ which is essentially independent of
$I_{rf}$.

Next we study the frequency dependence of $B_c$. Rather than
plotting $B_c$ versus $f$, we present the results by plotting the
frequency at which ML vanishes, denoted by $f_c$, versus $B$, as
done in Fig. \ref{fig3}(b). At low fields $f_c$ slowly increases
with field, but it starts to diverge at larger field. Such
divergence is expected when we identify $v_c=f_c a$ as the
ordering velocity proposed in \cite{Koshelevrecryst}. In the
present case, the plastic (or 'liquid') flow for $v < v_c$ is
controlled by the disordered VL's in the CE's, as we concluded
from simulations at small drive (as in
\cite{KokuboPRL02,BesselingPhDthesis}). This plastic flow involves
frequent slip events and transverse vortex jumps between rows.
Such behavior has also been observed in magnetic bubble arrays
moving along a rough wall \cite{Seshashear}. For $v>v_c$ these
effects disappear and coherent motion sets in. For $B\gtrsim 1.32$
T we could no longer resolve the ML effect below $200$ MHz. In
this field regime thermal fluctuations alone are sufficient to
induce incoherent motion, regardless of the dynamic influence of
disorder. The $f_c(B)$ data can approximately be fitted to $f_c
\sim (B_{m,e}-B)^{-\nu_B}$ with $\nu_B=2$ and an equilibrium
melting field $B_{m,e}=1.36$ T. An exponent $\nu_B=1$, as expected
from Eq.(\ref{DOvsTformula}) and the behavior of the melting line,
yields a rather poor fit. This discrepancy with the data may
originate from the fact that with changing field not only the
'distance' $B_{m,e}-B$ to the phase boundary varies, but also the
commensurability, i.e. the dislocation density and effective
disorder in the channel.

\begin{figure}
\epsfig{file=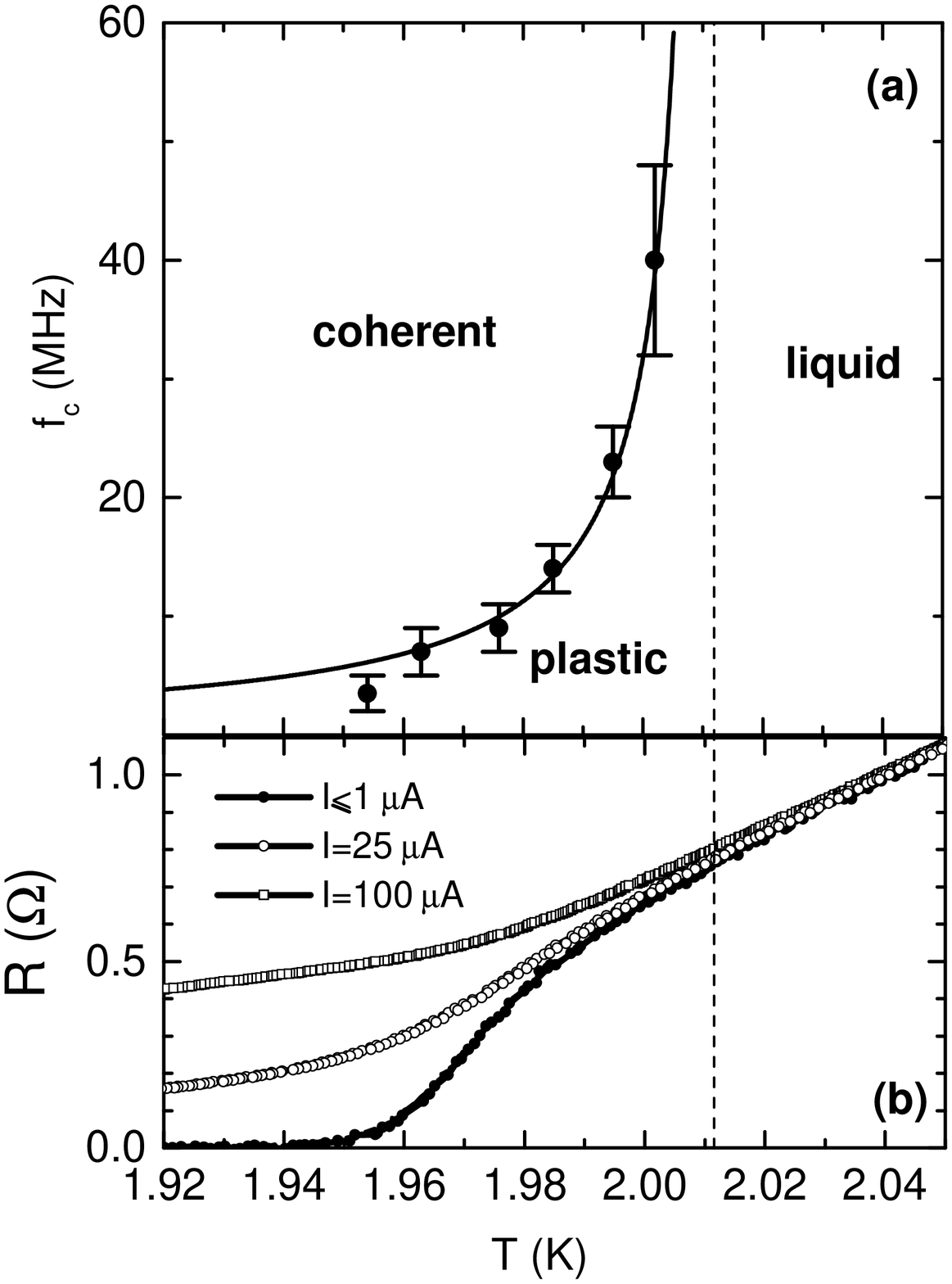, width=6.3cm} \vspace{0.2cm} \caption{(a)
Data points: dynamic ordering frequency at $B=1.16$ T versus
temperature. Drawn line: fit according to $f_c\sim
(T_{m,e}-T)^{-1}$ with $T_{m,e}=2.011$ K (indicated by the dashed
line). (b) Dc-resistance $V/I$ ($I_{rf}=0$) versus $T$ for several
dc-currents $I$.} \label{fig4}
\end{figure}

To avoid the possible influence of mismatch and provide a more
direct test of Eq.(\ref{DOvsTformula}), we determined $f_c$ as a
function of temperature for $B=1.16$ T (near matching) from an
analysis of the ML amplitude $\Delta I$ at several rf-currents
similar to that in Fig. \ref{fig3}(a) (details are given in
\cite{BesselingchanMLtobe}). The result is shown in Fig.
\ref{fig4}(a). Again a clear divergence of $f_c$ is observed. In
this case the data can be fitted quite well to
Eq.(\ref{DOvsTformula}) written as $f_c=f_0(1-T/T_{m,e})^{-1}$
with $f_0=0.174$ MHz and $T_{m,e}=2.011$ K. Interestingly, a
comparison with the {\it dc-resistance} at various currents (Fig.
\ref{fig4}(b)) shows that the equilibrium melting temperature
$T_{m,e}$ coincides with the temperature where the $R(T)$ curves
merge, i.e. the $IV$ curves become fully linear. This confirms the
usual assumption made in dc-transport studies on VL melting
\cite{Berghuis}. In measurements as function of field we observed
the merging of $R_{I\leq 1\mu A}(B)$ and $R_{I=100 \mu A}(B)$ at
$B\simeq 1.34$ T, in good agreement with $B_{m,e}$ determined
above. We further note that none of the dc-$IV$ curves show an
inflection point, implying that such feature is not required for
microscopic DO \cite{PaltielIV,DuartePRB96}.

It is remarkable that our system, in which 'edge disorder'
\cite{FeldmanPRL02_edgedisorder} is the prime source of pinning,
shows DO typical for a VL with {\em bulk disorder}. Let us
therefore discuss the nature of the DO in more detail. Even for a
2D VL with bulk pinning the DO was under intensive debate
\cite{GiaPRL96pluscom,Kolton99_02plusScheidl,BalentsdisdynPRB98,AransonplusScheidl}.
At present, the most likely scenario seems that, on increasing
$v$, after a crossover from fully plastic to partially layered,
smectic flow, finally a transition to a Moving Transverse Solid
occurs \cite{Kolton99_02plusScheidl}. At this {\em transverse
freezing transition} (TFT), inter-chain excursions (so called
permeation modes \cite{BalentsdisdynPRB98}) are suppressed, but
free dislocations with Burgers vector parallel to $\vec{v}$
\cite{AransonplusScheidl} and only longitudinal short-range order
(LSRO) remain. Particularly, it is the TFT which is described by
$T_{sh}$ \cite{Kolton99_02plusScheidl}. Turning to the channels,
we then suggest that our DO boundary $v_c(T,B)$ reflects the TFT
in which permeation modes due to roughness of the CE arrays are
suppressed. Preliminary simulations support this view
\cite{BesselingchanMLtobe}. Additionally, above $v_c$ we observe
only {\em incomplete} ML, suggesting indeed LSRO
\cite{BalentsdisdynPRB98} and residual slip between chains.

The shaking effect in the channels is estimated as follows. The
characteristic frequency $f_0=\Gamma_{p,v}/(k_BT_{m,e} a)$
obtained from Eq.(\ref{DOvsTformula}), can be derived from Ref.
\cite{Koshelevrecryst} as: $f_0=\sqrt{3/2\pi} \gamma_u
\rho_f/(\Phi_0^2a^2 d k_BT_ {m,e})$ with $\rho_f$ the flux flow
resistivity, $d$ the film thickness and $\gamma_u$ the pinning
energy squared times the 2D pinning range. For the channels, the
short wavelength ($\sim a_0$) disorder component due to vortex
displacements {\bf d} in the CE acts in a range $\sim a_0/2$ from
the CE's and has a strength $\sim A_s c_{66}$ with $A_s \sim
(\sqrt{\langle|{\bf d}|^2\rangle}/a_0)/(\pi \sqrt{3})$
\cite{BesselingPhDthesis,BesselingchanMLtobe}. We thus assume that
'shaking' of the first vortex layer near each CE dominates the
TFT. Hence, $\gamma_u \simeq (A_s c_{66}a_0b_0d)^2 (a_0/2)^2$.
Using the melting criterion $4\pi k_B T_{m,e}\simeq c_{66}a_0^2 d$
one obtains $f_0\simeq 20 A_s^2 \rho_f k_B T_{m,e}/\Phi_0^2 d$.
Taking $\rho_f \simeq \rho_n/2$ and $d=300$ nm yields $f_0=A_s^2
\cdot 500$ MHz. This is in reasonable agreement with the measured
value $f_0=0.174$ MHz when we assume $A_s \simeq 0.02$, i.e. rms
relative displacements in the CE of $(\sqrt{\langle|{\bf
d}|^2\rangle}/a_0)\simeq 0.1$.

In conclusion, dynamic melting of vortex matter driven through
disordered channels was studied by mode-locking experiments. The
melting line strongly depends on the ML frequency, i.e. the
average velocity. The associated ordering velocity diverges upon
approaching the equilibrium melting line, yielding a dynamic phase
diagram with coherent, plastic and fluid flow as predicted
theoretically \cite{Koshelevrecryst,Balents}. The ML technique
presents a powerful tool to study phase transitions in driven
periodic media and we hope our results will stimulate similar
investigations in related fields like CDW dynamics and solid
friction.

This work was supported by the Nederlandse Stichting voor
Fundamenteel Onderzoek der Materie (FOM).

\end{multicols}


\begin{references}

\bibitem{Blatterbible}G. Blatter {\it et al.}, Rev. Mod. Phys. {\bf 66}, 1125,
(1994).

\bibitem{Gruner}G. Gr\"uner, Rev. Mod Phys. {\bf 60}, 1129 (1988).

\bibitem{Perssonbook}B.N.J. Persson, {\it Sliding Friction:
Physical Principles and Applications} (Springer-Verlag, Berlin,
1998).

\bibitem{BhattaShi}S. Bhattacharya and M.J. Higgins, Phys.Rev. Lett. {\bf 70}, 2617
(1993); A.C. Shi and A.J. Berlinsky, Phys.\ Rev.\ Lett.\ {\bf 67},
1926 (1991).

\bibitem{Koshelevrecryst}A.E. Koshelev and V.M. Vinokur, Phys.\ Rev.\ Lett.\ {\bf 73}, 3580
(1994).

\bibitem{Jensen_filam_PRL88}H.J. Jensen {\it et al.},
Phys. Rev. Lett. \ {\bf 60}, 1676 (1988).

\bibitem{Balents}L. Balents and M.P.A Fisher, Phys. Rev. Lett. {\bf 75},
4270 (1995);

\bibitem{HellerGeers}M.C. Hellerqvist {\it et al.}, Phys.\ Rev.\ Lett.\ {\bf 76},
4022 (1996); Phys.\ Rev. \ B {\bf 56}, 5521 (1997); J.M.E. Geers
{\it et al.}, Phys.\ Rev.\ B {\bf 63}, 094511 (2001).

\bibitem{RyuPRL96}S. Ryu {\it et al.}, Phys.\ Rev.\ Lett.\ {\bf 77}, 5114 (1996);
M.C. Faleski {\it et al.}, Phys.\ Rev.\ B {\bf 54}, 12427 (1996).

\bibitem{Basslergradientsim} K. E. Bassler {\it et
al.}, Phys. Rev. B {\bf 64}, 224517 (2001).

\bibitem{PaltielIV}Y. Paltiel {\it et al.}, Phys. Rev. B {\bf
66}, 060503 (2002).

\bibitem{KokuboPRL02}N. Kokubo {\it et al.},  Phys.\ Rev.\ Lett.\ {\bf
88}, 247004 (2002).

\bibitem{Fiory}A. T. Fiory, Phys.\ Rev.\ B {\bf 7}, 1881 (1973);
A. Schmid and W. Hauger, J. Low. Temp. Phys. {\bf 11}, 667 (1973).

\bibitem{HarrisPRL95}J.M. Harris {\it et al.}, Phys. Rev. Lett.
{\bf 74}, 3684 (1995).

\bibitem{KoltonshapPRL01}A.B. Kolton {\it et al.}, Phys.\ Rev.\ lett.\ {\bf 86}, 4112 (2001).

\bibitem{fn}Contrary to usual $IV$ experiments, which can not
distinguish between changes in $v$ or in the fraction of moving
vortices, we directly obtain $v_c$ from the ML condition.

\bibitem{TheunissenPRL}M.H. Theunissen {\it et al.}, Phys. Rev. Lett. {\bf 77}, 159
(1996).

\bibitem{BesselingEPL03}R. Besseling {\it et al.}, Europhys. Lett. {\bf 62}, 419 (2003).

\bibitem{BesselingPhDthesis}R. Besseling, Ph. D. Thesis, Leiden University
(2001).

\bibitem{fnNbNpinfreq}The measurement frequency and dc/rf currents are much smaller than the
respective pinning frequency
$\omega_p/2\pi=(J_c\rho_f)^{NbN}/a_0B\gtrsim 5$ GHz ($\rho_f$ is
the flux flow resistivity) and pinning current $J_c\gtrsim 2\cdot
10^9$ A/m$^2$ of the NbN. The CE vortices can therefore be
considered as static.

\bibitem{ThorneMLampPRB87}R. E. Thorne {\it et al.}, Phys. Rev. B 35, 6360
(1987); Yu. I. Latyshev {\it et al.}, Phys. Rev. Lett. {\bf 87}
247007 2001.

\bibitem{Seshashear}R. Seshadri and R.M. Westervelt, Phys.\ Rev.\ Lett.\ {\bf 70}, 234 (1993);
Phys.\ Rev.\ B {\bf 47}, 8620 (1993).

\bibitem{BesselingchanMLtobe}R. Besseling {\it et al.}, in
preparation. See also N. Kokubo {\it et al.}, cond-mat/0308512.

\bibitem{Berghuis}P. Berghuis {\it et al.}, Phys. Rev. Lett. {\bf 65}, 2583
(1990); Phys.\ Rev.\ B {\bf 47}, 262 (1993).

\bibitem{DuartePRB96}A. Duarte {\it et al.}, Phys. Rev. B {\bf 53}, 11336
(1996).

\bibitem{FeldmanPRL02_edgedisorder}D.E. Feldman and V.M. Vinokur, Phys. Rev. Lett. {\bf 89},
227204 (2002).

\bibitem{GiaPRL96pluscom}T. Giamarchi and P. Le Doussal, Phys. Rev. Lett. {\bf 76}, 3408
(1996); L. Balents {\it et al.}, Phys. Rev. Lett. {\bf 78}, 751
(1997); P. Le Doussal and T. Giamarchi, Phys.\ Rev.\ B {\bf 57},
11356 (1998).

\bibitem{Kolton99_02plusScheidl}A.B. Kolton {\it et al.}, Phys.\ Rev.\ Lett.\ {\bf 83}, 3061
(1999); Phys.\ Rev.\ Lett.\ {\bf 89}, 227001 (2002); S. Scheidl
and V.M. Vinokur, Phys.\ Rev.\ B {\bf57}, 13800 (1998).

\bibitem{BalentsdisdynPRB98}L. Balents {\it et al.}, Phys. Rev. B {\bf 57}, 7705 (1998).

\bibitem{AransonplusScheidl}I.S. Aranson {\it et al.}, Phys. Rev. B {\bf 58},
14541 (1998).

\end{references}
\end{document}